\documentclass{article}

\usepackage{geometry}
\usepackage{doc}

\usepackage{url}

\usepackage{wrapfig}
\usepackage{xcolor}
\usepackage{colortbl}
\usepackage{amsmath}
\usepackage{amsfonts}
\usepackage{paralist}
\usepackage[sort&compress, numbers]{natbib}
\usepackage{url}
\usepackage{color}
\usepackage{xspace}
\usepackage{caption}
\usepackage{subcaption}
\usepackage[shortlabels]{enumitem}
\usepackage[colorlinks=true,linkcolor=blue,citecolor=red]{hyperref}
\usepackage{wrapfig}

\usepackage{graphicx}
\usepackage{epstopdf}
\usepackage{geometry}
\newcommand{\wilee}{WILEE\xspace}
\newcommand{\wileeGPLong}{Genetic Perturbation Engine\xspace}
\newcommand{\wileeGP}{GPE\xspace}
\newcommand{\malmo}{Malmo\xspace}
\newcommand{\Malmo}{Malmo\xspace}
\newcommand{\IOCLong}{\textit{indicator of compromise}\xspace}
\newcommand{\IOC}{\textit{IOC}\xspace}

\title{\vspace{-1.0cm}Automating Cyber Threat Hunting Using NLP, Automated Query Generation, and Genetic Perturbation}

\makeatletter
\renewcommand\@date{{%
  \vspace{-\baselineskip}%
  \large\centering
  \begin{tabular}{@{}c@{}}
    Prakruthi Karuna\textsuperscript{1} \\
    \small pkaruna@perspectalabs.com  \hspace{0.5 cm}
  \end{tabular}%
  \begin{tabular}{@{}c@{}}
    Erik Hemberg\textsuperscript{2} \\
    \small hembergerik@csail.mit.edu
  \end{tabular}
  \begin{tabular}{@{}c@{}}
    Una-May O'Reilly\textsuperscript{2} \\
    \small unamay@csail.mit.edu
  \end{tabular}%
  \begin{tabular}{@{}c@{}}
    Nick Rutar\textsuperscript{1} \\
    \small nrutar@perspectalabs.com
  \end{tabular}

  \textsuperscript{1}Perspecta Labs, Inc.
  \textsuperscript{2} MIT CSAIL
}}
\makeatother

\begin{document}

\twocolumn[
  \begin{@twocolumnfalse}
  \maketitle
  \end{@twocolumnfalse}
]

%
%
\section{Introduction}
\label{introduction}

Scaling the cyber hunt problem poses several key technical challenges. Detecting and characterizing cyber threats at scale in large enterprise networks is hard because of the vast quantity and complexity of the data that must be analyzed as adversaries deploy varied and evolving tactics to accomplish their goals. There is a great need to automate all aspects, and, indeed, the workflow of cyber hunting. AI offers many ways to support this.  We have developed the  \wilee system that automates cyber threat hunting by translating high-level threat descriptions into many possible concrete implementations.  Both the (high-level) abstract and (low-level) concrete implementations are represented using a custom domain specific language (DSL). \wilee uses the implementations along with other logic, also written in the DSL, to automatically generate queries to confirm (or refute) any hypotheses tied to the potential adversarial workflows represented at various layers of abstraction. We summarize \wilee in Section~\ref{sec:wilee}.

\begin{figure}[t]
\includegraphics[width=3.35in]{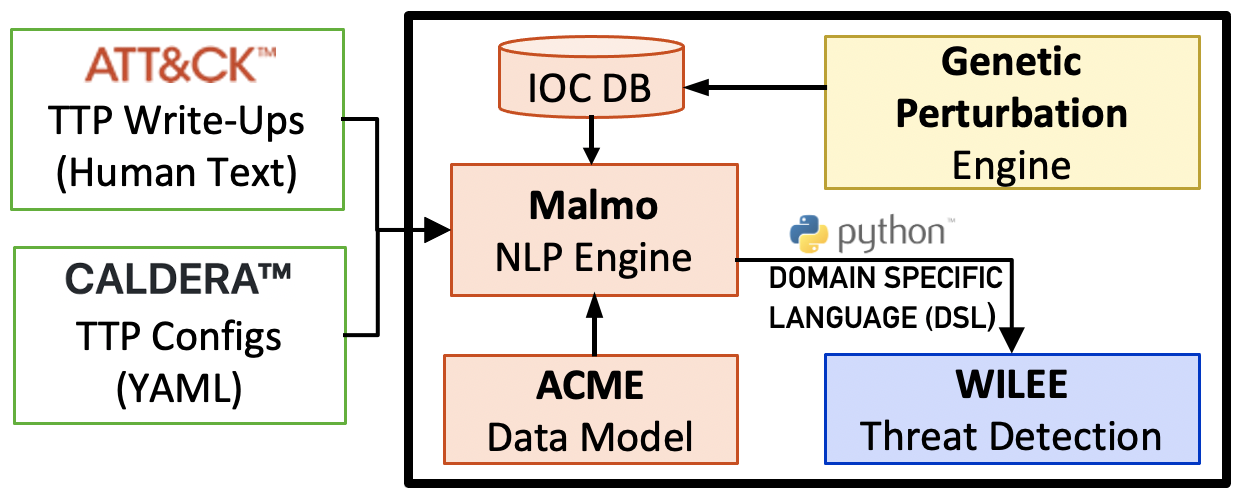} 
\centering
\caption{AI Components for automated DSL Generation for use within the WILEE threat detection system.}
\label{fig:overview}
\end{figure}

In this contribution we focus on presenting two AI components of \wilee that coordinate to support its  automated query generation for hunting, see Figure~\ref{fig:overview}. First, in Section~\ref{sec:malmo}, we introduce a hunt component, named \malmo, that uses Natural Language Processing (NLP) to automate extraction and translation of known threat descriptions.  Known threats can be identified with pattern recognition and signature-based methods. Their descriptions are embedded within human-readable, semi-structured documents. \Malmo extracts and translates, to machine digestible format, key fields of the documents including indicators of compromise, to make them operational for the hunt.  

Known threats evolve into unknown threats. Evolutionary adaptations occur at the level of TTPs, i.e. at the behavioral level,  or, at the lower level of simple, but effective  modifications of their signatures. This implies that, to detect unknown threats, explorative variations should also be hunted. The hunt needs to be widened, while not becoming so wide as to overwhelm data collection and its subsequent filtering. To address these challenges,  \wilee  introduces a genetic programming hunt component~\cite{mcdermott2015genetic}. Called \wileeGPLong, (\wileeGP{}),  the component is integrated both within \wilee{'s} workflow, as a  generator of perturbed threat implementation, see Figure~\ref{fig:wilee_gp}, and with \malmo, see Figure~\ref{fig:genperturbAndMalmo} as a way to replace a field describing an \IOCLong (\IOC{}) of a known threat, with an alternative of the same type, drawn from a database of possibilities.  We summarize \wileeGP in Section~\ref{sec:wileeGP}.

Conclusions, and primarily future work directions are presented in Section~\ref{sec:concl-future-work}.
\let\thefootnote\relax\footnotetext{This material is based upon work supported by the DARPA Advanced Research Project Agency (DARPA) and Space and Naval Warfare Systems Center, Pacific (SSC Pacific) under Contract No. N66001-18-C-4036. The views, opinions, and/or findings expressed are those of the author(s) and should not be interpreted as representing the official views or policies of the Department of Defense or the U.S. Government.  \textbf{Approved for Public Release, Distribution Unlimited.}}

\section{WILEE Framework and DSL}
\label{sec:wilee}

\begin{figure}
\includegraphics[width=3.3in]{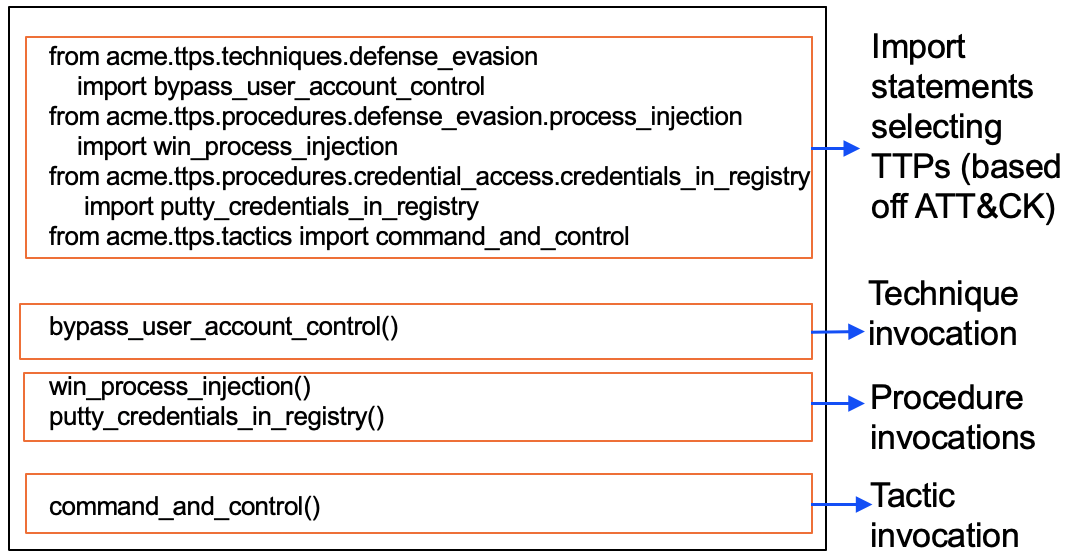} 
\centering
\caption{(Abstract) Threat Description DSL Example}
\label{threat-description}
\end{figure}

The primary input to the WILEE threat detector is a \textit{\textbf{Threat Description}}, which is expressed in the DSL.   This initial input is a high level, mostly abstract, workflow that the detector is hunting.  Each `call' from this description is an abstract TTP (tactic, technique, or procedure) mapping primarily to MITRE ATT\&CK\textregistered   \cite{mitreattack} entries.   An example of this input is shown in Figure \ref{threat-description}.   This input is technically optional, and if an abstract workflow is not defined, WILEE will attempt to detect the full `kill-chain'\cite{killchain}. 

\begin{figure}
\includegraphics[width=3.3in]{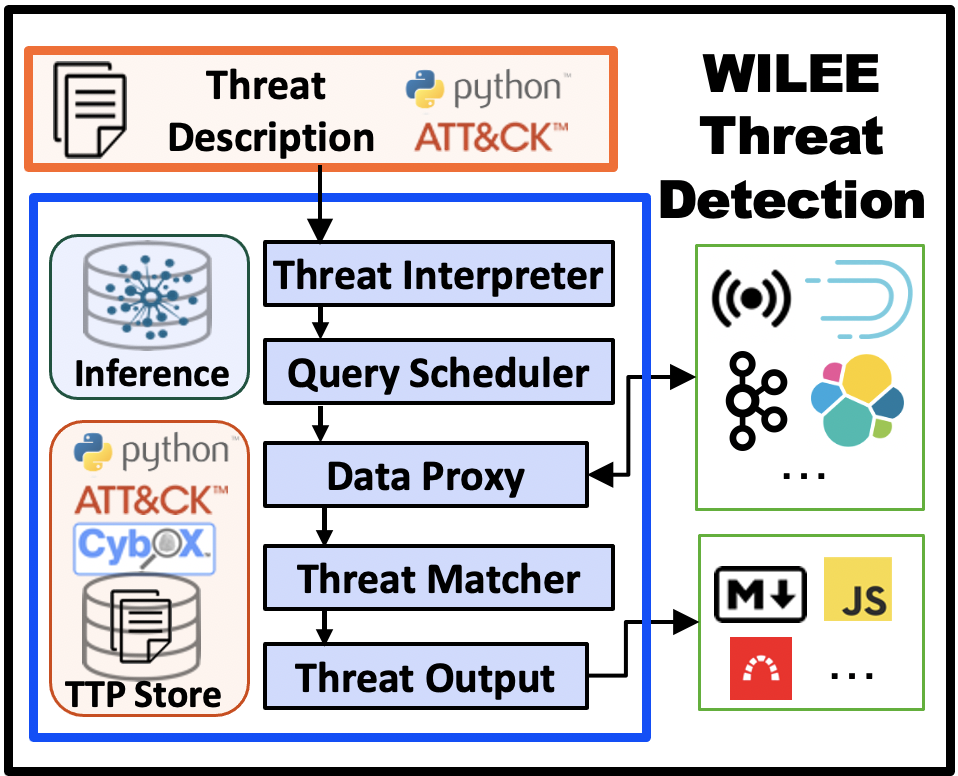} 
\centering
\caption{WILEE Threat Detector Framework}
\label{wilee-threat}
\end{figure}

\begin{figure}
\includegraphics[width=3.3in]{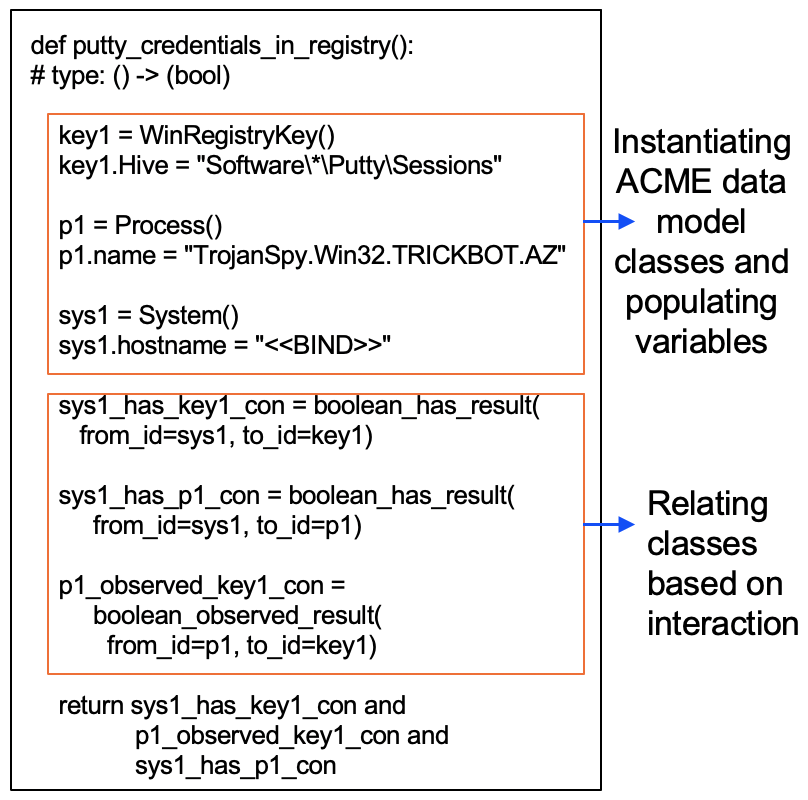} 
\centering
\caption{(Concrete)  TTP DSL Example}
\label{ttp-dsl}
\end{figure}

The \textit{\textbf{Threat Interpreter}} takes in the Python DSL (if explicitly defined) and pulls all appropriate tactics, techniques, and procedures from the  \textit{\textbf{TTP Store}} to concretize all possible combinations of the abstract workflow.  An example of one TTP representation is shown in Figure \ref{ttp-dsl}.  While the DSL is Pythonic in syntax, that code is never executed.  Rather the Python AST is parsed to generate all the possible queries needed to query/inquire whether the events in the workflow are occurring on the protected enclave.   This generation of the queries occurs in the  \textit{\textbf{Query Scheduler}}, which is data store agnostic.  The actual queries for the supported data stores takes place in the \textit{\textbf{Data Proxy}} and the query results allow WILEE to ingest the confirmed (or refuted) hypotheses as a graph with TTP-based inferences between entities, into an internal data store.   A match algorithm is applied to the edges within that data store in the  \textit{\textbf{Threat Matcher}} module.  Finally, the results are output to various formats (including Markdown, JavaScript, and ticketing systems like Redmine) in the \textit{\textbf{Threat Output}} module.

The WILEE framework has been successfully deployed to categorize threats using data sets including the Operationally Transparent Cyber (OpTC) evaluation set\cite{opttc} and in categorizing MITRE CALDERA \cite{caldera} based simulated attacks.   In both of these deployments, the DSLs representing the individual TTPs that were used for categorization, and maintained/stored in the WILEE TTP Store, were manually generated by a subject matter expert (SME).    While the WILEE system is fully automated, the dependence on these manually generated entries limits the breadth of TTPs WILEE can detect, and lessens the speed at which it can model new TTPs as they are incorporated within ATT\&CK and other threat models.  The rest of this paper will discuss how we are automating the process of DSL generation utilizing NLP within Malmo and novel IOC generation within GPE.

\section{Malmo: Natural Language Processing (NLP) Engine}
\label{sec:malmo}

\begin{figure}
\includegraphics[width=3.35in]{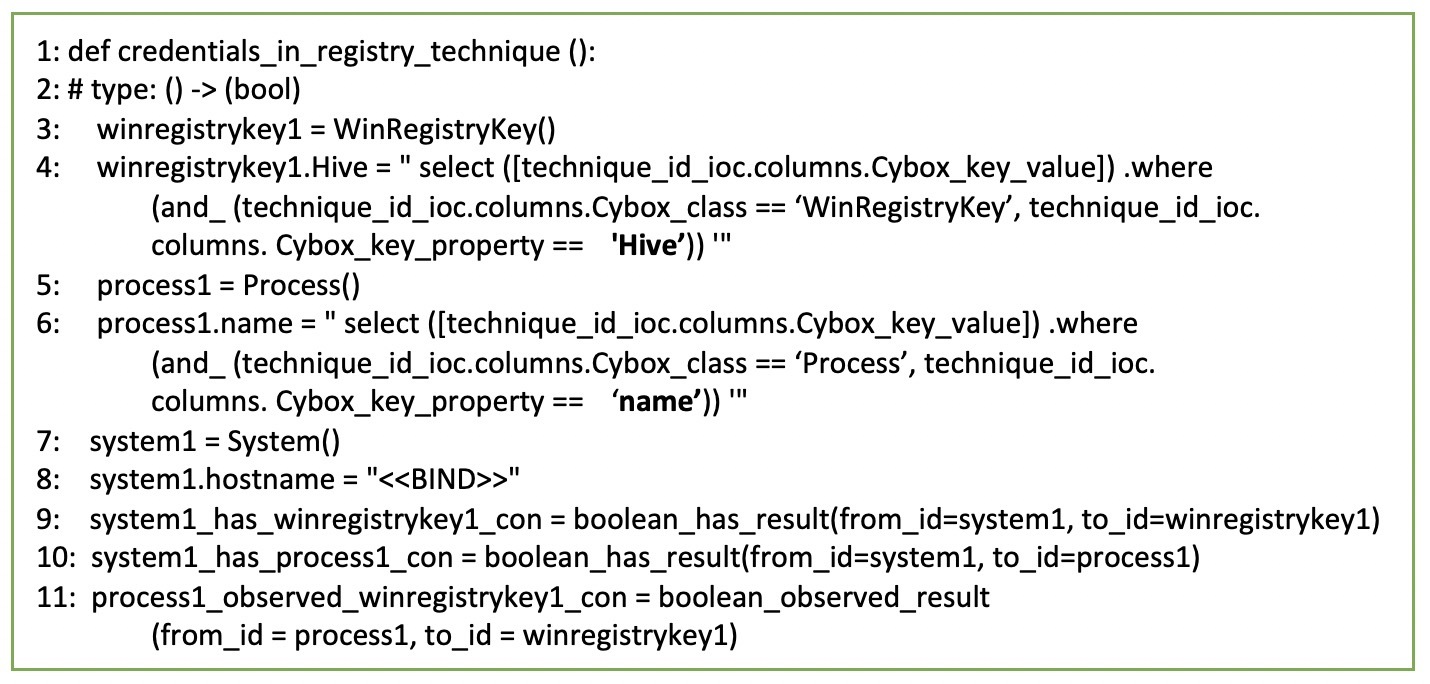} 
\centering
\caption{(Concrete)  Malmo DSL Example}
\label{malmo-dsl}
\end{figure}

MITRE ATT\&CK provides a knowledge base of adversarial tactics, techniques and procedures (TTPs) written in a human readable format. However, it's challenging for a computer to understand or hunt based on this knowledge base as it's not easily digestible for a cyber hunt system.  MITRE ATT\&CK TTPs refer to several network and host level entities such as "system", "window registry keys", "process". It is necessary to model these entities to be able to capture the knowledge in the TTPs, therefore, we use MITRE CybOX (Cyber Observable Expression) as our base data model. However, CybOX does not contain all the entities referred to in TTPs, therefore we extend this data model to include objects referred to in the TTPs  (e.g, active directory in windows OS) but are not modeled in CybOX. Once we have a detailed data model, we use this to write our pythonic DSL that represents TTPs in a computer digestible format. In our DSL we represent each TTP technique as a function as shown in Figure \ref{ttp-dsl}. Within each function, we instantiate objects from our data model and relate them using relations of type: “has” and “observed”. By doing so we can represent all the essential objects referred to in the technique and also relate them to form meaningful triplets of form:  “system – has – process”, process – observed – win\_registry\_key”. We populate the variables associated with our objects to values that are relevant to the technique. For example Figure \ref{ttp-dsl} shows the DSL for ATT\&CK technique \textit{T1552.002: Unsecured Credentials: Credentials in Registry} where we populate variables winregistrykey1.Hive to "Software\textbackslash *\textbackslash Putty\textbackslash Sessions" that is derived from the technique description and variable process1.name to "TrojanSpy.Win32.TRICKBOT.AZ ". By doing this we are tying  indicators of compromise (IOC) to different variables in our TTP description represented using the DSL. Indicators of compromise are forensic artifacts that are used as signs that a system has been attacked or worse compromised by an attack or that it has been infected with a particular malicious software. In this work, we use IOCs to identify attacker activity. Finding data points using IOCs informs us with high confidence that there was attacker activity. However, each variable could be associated with multiple IOC values. Therefore, we store these IOCs in a database and query it on the fly. To do this we populate object variables with sql queries written using SQLAlchemy notations as shown in Figure \ref{malmo-dsl}.

We can also model red team tool-based indicators using our DSL. We worked with the red team tool CALDERA's stockpile plugin that provides a set of commands to replicate adversary behaviors as if a real intrusion is occurring. An example command is: 
"Get-Process -Name "powershell" | Stop-Process"
where an attacker abuses command and script interpreter as described in technique \textit{T1059.001}. We use such commands to identify attacker activity.

\subsection{Automated DSL Generation}
\begin{figure}
\includegraphics[width=3.35in]{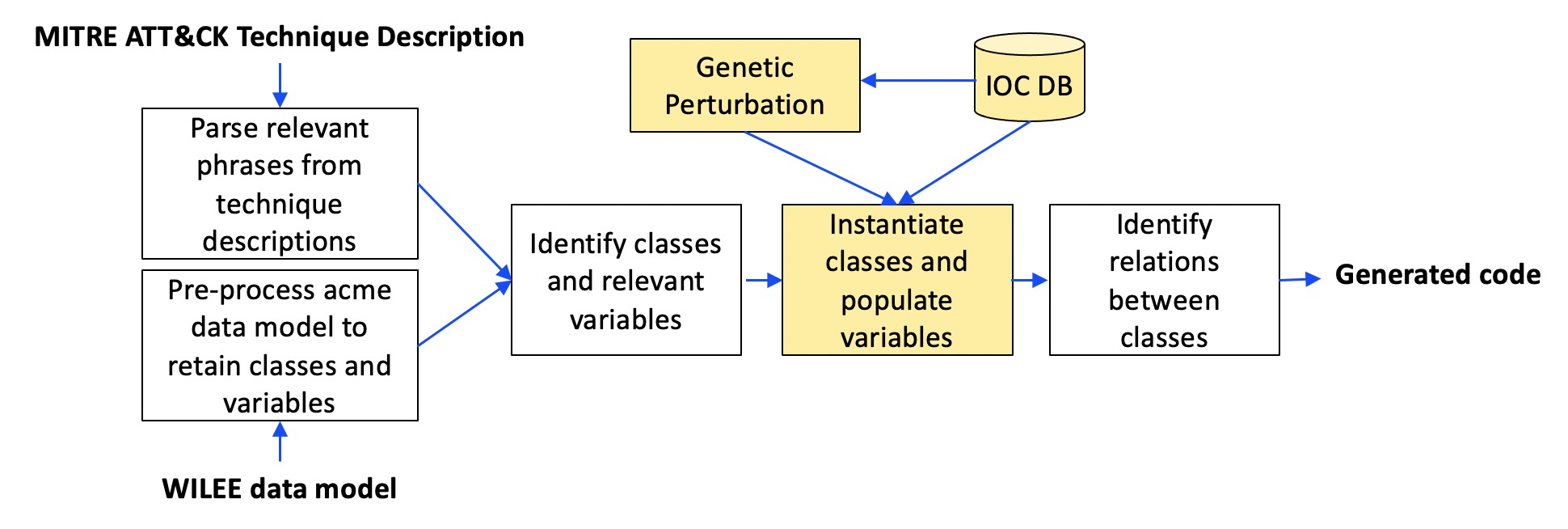} 
\centering
\caption{Pipeline for DSL generation}
\label{malmo-pipeline}
\end{figure}

Threats are ever evolving, and new techniques and procedures get added to the MITRE ATT\&CK repository often. However, generating such a detailed DSL is a labor-intensive task. Therefore we use NLP techniques to automate the generation of pythonic DSL from MITRE ATT\&CK TTPs. Our NLP pipeline is as shown in Figure \ref{malmo-pipeline}.

We take as input TTP technique descriptions, our data model, and indicators of attacker activity. We parse relevant phrases from technique description by extracting noun phrases that follow the regular expression pattern of Adjective*Noun+. We then pre-process our data model to retain classes and variables. We then match the classes and variables to the phrases from text to identify relevant classes and variables. To identify relevant classes we compute the relevancy of a class to a TTP technique based on the formulas below and include top $n$ number of classes with the highest class inclusion value. We then instantiate classes and populate variables with indicators of attacker activity using template filling techniques. We finally relate these classes based on previously seen relations.  

 \[
\begin{array}{c}
\exists \textrm{ Word match if }  word_{noun\_phrase} \, \in \, class\_name \, \\
\textrm{ and } word_{noun\_phrase} \in class.variable\_name \\

\\

\textrm{Relative importance of Word match} = \\
\frac{1}{(\textrm{Frequency of }  word_{noun\_phrase} \textrm{ in data model)}} \\

\\

\textrm{Percentage words matched in Word match} = \\
\frac{\textrm{Number of matched } word_{noun\_phrase}}
{\textrm{Number of } words_{variable\_name}} * 100 \\

\\

\textrm{Word match value} = (\textrm{Relative imp of Word match})  * \\
 (\textrm{Percentage words matched in Word match}) \\

\\

\textrm{Class inclusion value} =  \\
\sum_{variable \in class} \textrm{Word\ match value} \\
\end{array}
\]

\newpage

\section{Genetic Perturbations of IOCs}\label{sec:wileeGP}

 
\wilee uses genetic programming (GP) as the algorithm of \wileeGP. 
Genetic Programming is an evolutionary algorithm where candidate solutions are executable code that are manipulated in their parse tree representation. GP's genetic operators can exchange the subtrees of code trees and maintain their ability to execute without syntactic errors.  A code tree is executed and its compliance with output requirements determines its fitness. We use grammar-guided GP to allow the search space of GP to be defined independently of its variation operators.  For more information see~\cite{mcdermott2015genetic,poli2008field,o2019introduction,mckay2010grammar}. 
We chose GP for \wilee because abstract syntax trees (ASTs) can be extracted from \wilee's threat implementations, as they are represented as Pythonic DSL.  With parse tree representations, they can be manipulated by GP within the \wileeGP to vary the hunt queries and \IOC{s}. This allows search at the abstract threat level.

\begin{figure}[h]
  \centering
  \includegraphics[width=0.5\textwidth]{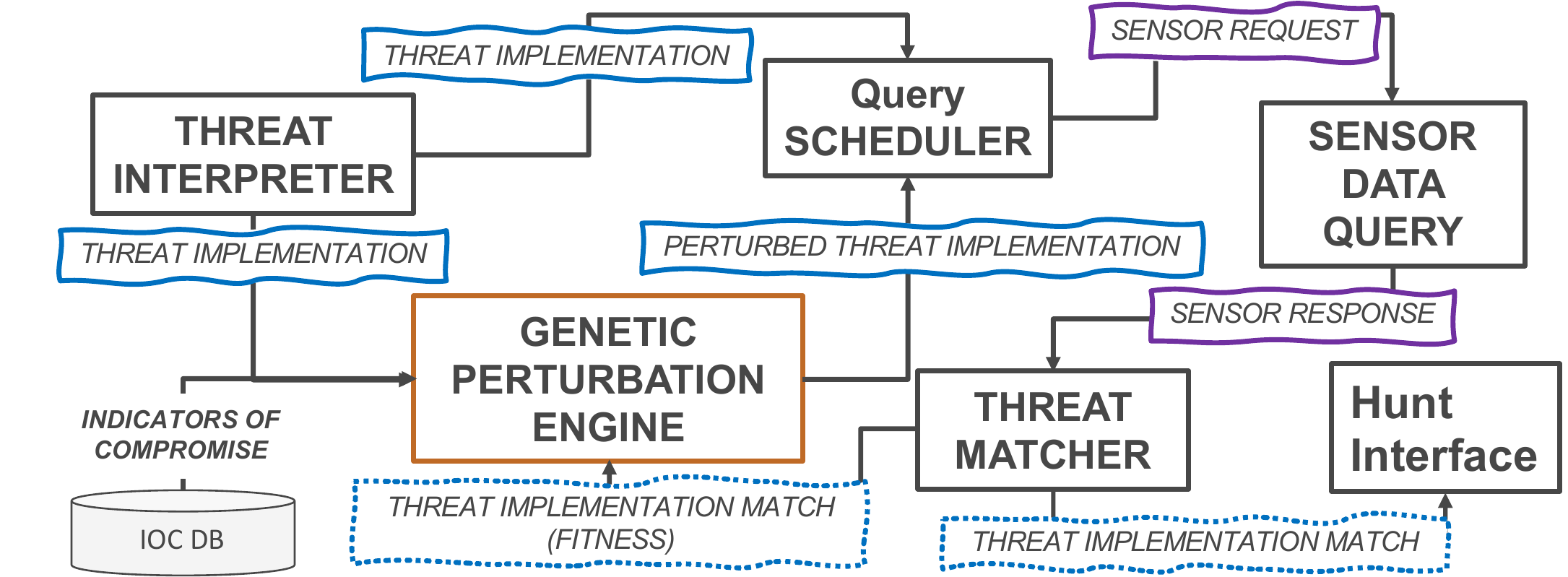}
  \caption{\wileeGP intercepts \texttt{Threat Implementations} and perturbs them.}
  \label{fig:wilee_gp}
\end{figure}

\newcommand{\TI}{\texttt{Threat Implementation}\xspace}

\noindent \textbf{Threat Implementation Perturbation:} Per  Figure~\ref{fig:wilee_gp}, \wileeGP and its \texttt{Threat Implementation} perturbation role are best described in the
context of the standard \wilee workflow.    
\wileeGP is an optional injection into this flow. It is fed a diverted \texttt{Threat Implementation} from the
\texttt{Threat Interpreter} and then perturbs it to enlarge the space of
variants around the \texttt{Threat Implementation},  using adaptations that are GP mutations or type-matched options from the \IOC database that is the cornerstone of \malmo. 
We encounter several critical challenges. The first is that the GP system starts without fitness values for one or more candidate \TI{s} because their match quality is unknown.  We solve this challenge by using novelty search; a GP technique where genetic selection prefers candidates that are different from each other, \cite{hemberg2019domain,lehman2010efficiently}. We have improved novelty search to provide it with a knob-like feature that automatically balances each population between novel membership and fitness-based membership.\cite{kelly2019improving} The second (open) challenge is that the fitness of a \TI at one point in time may be inaccurate at another. For example, a \TI could be explored before a C2C domain is established and thus be unfit. But, in a system where, some time later, a C2C is established, it would be fit.  The fitness of a \TI can also change if the filtering of a data stream changes and, with the change its match changes.  

\begin{figure}[!tb]
  \centering
  \includegraphics[width=0.25\textwidth,height=1.25in]{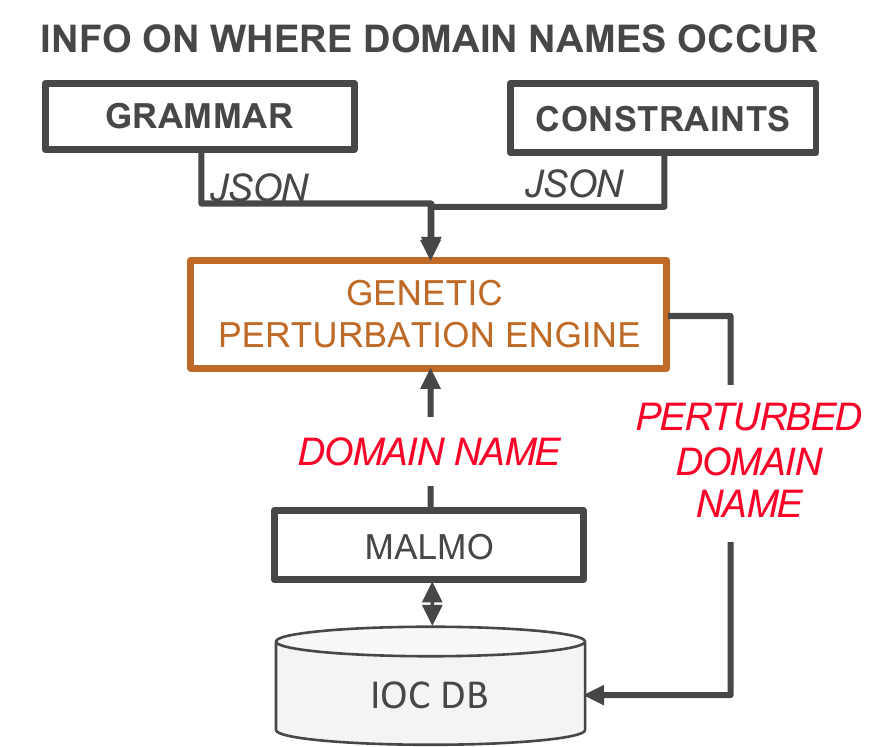}
  \caption{MALMO and GP creating new IOCs.}
  \label{fig:genperturbAndMalmo}
\end{figure}

\noindent \textbf{IOC Perturbation:} Figure~\ref{fig:genperturbAndMalmo} shows a second use of \wileeGP.  We have implemented the ability for \wileeGP to look for $<$\texttt{bind}$>$ values in the \TI, and, on the basis of the type required, access matching options in \malmo{'s} IOC database.  This allows the substitution of known threat information into concretized \TI{s}.  We exploit \wileeGP's aforementioned grammar, some constraint logic, and (currently) blind mutation to do so.  The ultimate use case we have in mind would perturb bound values with novel values that are informed by domain information.  For example, the Hive locator in ATT\&CK technique t1552.002 which is \texttt{Software\slash SimonTatham\slash Putty\slash Sessions} could be intelligently cycled through reasonable options that replace the putty program author's name.  While we have not assigned fitness to various perturbations, this future work may need to be handled with our novelty search.

%

\section{Conclusion and Future Work}
\label{sec:concl-future-work}
We have presented AI-based techniques for the automation of DSL entries in support of threat detection.  The primary focus for future work will be the evaluation of these techniques.  We currently have 'ground truth' for which TTPs are deployed in adversarial simulations in systems like CALDERA.  Our immediate next step is evaluating how WILEE performs when using SME generated DSLs in comparison to automatically generated DSLs.  For next steps for IOCs, when using IOC based search data, we can learn attacker behaviors and later use these learnt behavior to identify new IOCs. In the future, we plan to identify variations of CALDERA commands to further identify threats. Finally, we are considering the value of the \wileeGP adding perturbations that
are like regular expressions~\cite{michael2019}, directing the hunt to a family of variants to which a threat may evolve.

\newpage
\bibliography{bibliography}
\bibliographystyle{unsrt}

\end{document}